\documentclass[dvips]{article}
\usepackage{icrctc07}

\title{First simultaneous multiwavelength observations of the BL Lac object 1ES1959+650 in a steady state with MAGIC and Suzaku/Swift}

\shorttitle{MWL observations for 1ES1959+650 with MAGIC and Suzaku/Swift}
\authors{M.~Hayashida$^{1}$, C.~Bigongiari$^2$, D.~Kranich$^3$, A.~Moralejo$^{4}$  for the MAGIC collaboration.\\
G.~Tagliaferri$^5$, F.~Tavecchio$^5$, L. Foschini$^6$ , G.~Ghisellini$^5$, L.~Maraschi$^7$, G.~Tosti$^8$
}

\shortauthors{M. Hayashida et al}
\afiliations{$^1$Max-Planck-Institut f\"ur Physik, M\"unchen, Germany, 
$^2$Universita di Padova and INFN, Padova, Italy
$^3$ETH Zurich,Switzerland, 
$^4$Institut de F\'\i sica d'Altes Energies, Edifici Cn., Barcelona, Spain,
$^5$INAF/Osservatorio Astronomico di Brera, Merate, Italy, $^6$INAF/IASF, Bologna, Italy, 
$^7$INAF/Osservatorio Astronomico di Brera, Milano, Italy,  $^8$University of Perugia and INFN, Perugia, Italy
}
\email{mahaya@mppmu.mpg.de}

\abstract{Simultaneous multiwavelength observations were conducted for the BL Lac object 1ES1959+650 in a steady state in May 2006 with the MAGIC telescope and the X-ray satellites Suzaku and Swift. Swift can also provide multi-filter photometry in the UV-optical band. The source was clearly detected in all observed energy bands, from the optical to TeV. With respect to previous observations the source was in a low state in the very high energy (VHE) band ($\sim$10\% Crab flux above 300 GeV) but in a relatively high state in the X-ray band. The light curves showed rather stable activities, with no significant variability in the VHE $\gamma$-ray emission and small variability ($\sim$ 10\% amplitude) in the X-ray band. The observed spectral energy distribution in the steady state can be described by a one-zone synchrotron self-Compton model.}

\begin{document}
\maketitle

\section{Introduction}

Multiwavelength observations of the BL Lac objects can provide information on the acceleration mechanism and the properties of the jet. Since the BL Lac objects are highly variable sources, simultaneous multiwavelength observations are essential tools to test the various emission models.
At present, it is unclear whether the observed emission is entirely due to accelerated electrons, as in synchrotron self-Compton (SSC) models~[e.g.]\cite{CandC}, or whether the high-energy emission is due to secondary electrons produced by accelerated protons and ions~[e.g.]\cite{Mue01}.

1ES1959+650 (RA\,=\,19:59:59.95,\,Dec\,=\,+65:08:54 [J2000], $z$\,=\,0.047) is categorized as a High-frequency BL Lac object (HBL) where the synchrotron peak lies in the X-ray regime~\cite{Tag03}. The first TeV $\gamma$-ray signal from this object was reported in 1998~\cite{Nis99}. Recent results from the observations in 2004 with the MAGIC telescope show a signal at the 8~$\sigma$ level and a flux of about 20\% Crab ($>$ 300 GeV) from 6.5 hours of observation~\cite{MAGIC1959}. This source is particularly interesting for multiwavelength observations because a so-called "orphan" flare in TeV band without any X-ray counterparts was observed in 2002. This activity can not be explained by conventional one-zone SSC models~\cite{Kra04}.


\section{Observations}
\subsection{MAGIC Observations}
The MAGIC (Major Atmospheric Gamma Imaging Cherenkov) telescope is an Imaging Atmospheric Cherenkov Telescope (IACT) with a 17-m diameter dish, located on the Canary Island of La Palma (28.2$^{\circ}$~N, 17.8$^{\circ}$~W, 2225 m\,a.s.l.).  The telescope is operating at a $\gamma$-ray trigger threshold of $\sim50$\,GeV and a spectral threshold of $\sim$100\,GeV. The telescope parameters and performance are described in detail in~\cite{Crab}.

 \begin{figure}[ht]
  \centering
   \includegraphics[width=7cm, clip]{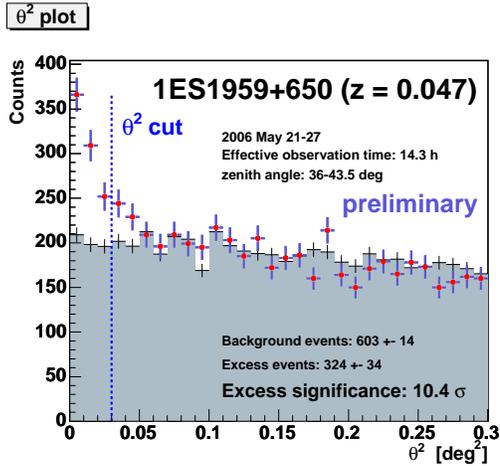}
  \caption{$\theta^2$ distribution of 1ES1959+650 above 350 photoelectron (corresponding to the $\gamma$-ray peak energy of $\sim$400 GeV) . A cut of $\theta^2 < 0.03\,{\rm deg^2}$ (dashed blue line) yields the signal of 324 events over 603 normalized background events, corresponding to a 10.4 $\sigma$ excess.
   }
   \label{SkyMap}
   \end{figure}

%

1ES1959+650 was observed with the MAGIC telescope from 21st to 27th of May, 2006.
The zenith angle during the observations was in the range from $36^{\circ}$ to $43.5^{\circ}$.
Observations were performed in wobble mode~\cite{Dau97} with the object offset by $0.4^{\circ}$ from the camera center.
After the quality selection of the data the total effective observation time was 14.3 hours.
The analysis was performed using the standard MAGIC analysis software. Details of the analysis chain can be seen in~\cite{Crab}.
In order to determine the significance of the $\gamma$-ray signal and the number of excess events,  a final cut with $\theta^2$ parameter was used. The $\theta$ is the angular distance between the source position in the sky and the reconstructed arrival position of the air shower which is estimated using "DISP" method~\cite{DISP}. A cut of $\theta^2 < 0.03\,{\rm deg^2}$ yields an excess of 324 events over 603 normalized background events, corresponding to a 10.4 $\sigma$ excess as shown in figure~\ref{SkyMap}.
Only events $>$ 350 photoelectrons were used, corresponding to the peak of the energy distribution at about 400 GeV for $\gamma$-ray. 

\subsection{Suzaku Observations}
Suzaku~\cite{Mit07} is a X-ray satellite with four X-ray Imaging Spectrometers (XIS) and a separate Hard X-ray Detector (HXD). The XIS are sensitive in the 0.2-10 keV band with two types of CCDs composed of front-illuminated CCDs (for XIS0, 2 and 3) and a back-illuminated type (XIS1). XIS1 is more sensitive below 2 keV. The HXD is a silicon PIN diode array and is the most sensitive detector in the 10-70 keV band thanks to the good noise shielding.

The Suzaku satellite was pointing 1ES1959+650 between 01:13:23 of May 23rd and 04:07:24 of May 25th, 2006 (UT).
The HXD/PIN light curve shows a rapid increase of the noise after about 100 ksec (possibly due to the in-orbit unexpected radiation~\footnote{see http://www.astro.isas.ac.jp/suzaku/log/hxd/}) and the data after this event cannot be used for the analysis. After screening the total net exposure time for XIS and HXD are 99.3 ksec and 40.2 ksec, respectively.
According to the RXTE/ASM light curve of May 2006~\footnote{see http://heasarc.gsfc.nasa.gov/xte\_weather/}, the source was in a relatively high state during the Suzaku pointing.

\subsection{Swift Observations}
The Swift satellite~\cite{[4]}, with its easy and flexible scheduling, can be optimally employed for the observation of bright blazars. It carries three instruments. The Burst Alert Telescope (BAT) is optimized for 15-150  keV  with high sensitivity while the X-Ray Telescope (XRT) is sensitive in the 0.3 -10 keV band. In addition a UV/optical telescope can provide data in the 170-600 nm band.
The source was also observed with Swift around the Suzaku pointing 19th, 21st and from 23rd to 29th of May.

\section{Results and Discussions}

\subsection{Light curves}

\begin{figure}[htbp]
  \centering
 \includegraphics[width=7cm, clip]{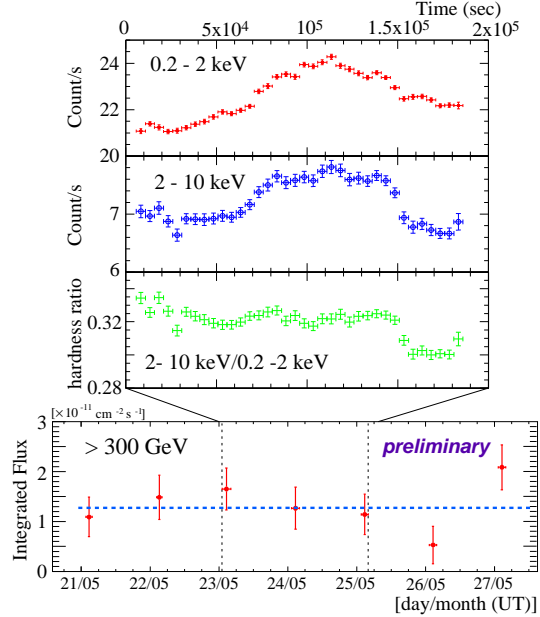}
   \caption{Light curves of 1ES1959+650 during this campaign. \textbf{[Upper]:} X-ray light curves obtained by the Suzaku XIS1 detector. \textit{Top to bottom:} Count rates in the 0.2-2 keV band, in the 2-10 keV band and the hardness ratio (2-10 keV/0.2-1 keV). \textbf{[Lower]:} Diurnal light curve of VHE $\gamma$-ray ($>$ 300 GeV) measured by the MAGIC telescope. A dotted horizontal line represents the average flux. Dotted vertical lines show the observation window of the Suzaku pointing.
   }
   \label{MWLC}
   \end{figure}

The light curves in VHE $\gamma$-ray and  X-ray are shown in figure~\ref{MWLC}. The average integrated flux above 300 GeV is $(1.27 \pm 0.16)\times 10^{-11} {\rm cm}^{-2} {\rm s}^{-1}$ ($\chi^2$/d.o.f = 8.5/6), which corresponds to about 10\% of the Crab Nebula flux~\cite{Crab}. No significant strong variability can be seen in VHE $\gamma$-ray emission.
In the X-ray band a flare of small amplitude ($\sim$ 10\%) with a rising time of $T_{\rm r} \sim 10^5$ s could be observed. The hardness ratio shows rather small spectral variation, except for a the sudden drop at $t \sim 1.5 \times 10^5$s.

\subsection{Spectra}
Since the source was stable both in the VHE $\gamma$-ray and the X-ray bands we extracted the spectra using the whole dataset during the campaign for each band.

\begin{figure}[htbp]
  \centering
 \includegraphics[width=7cm, clip]{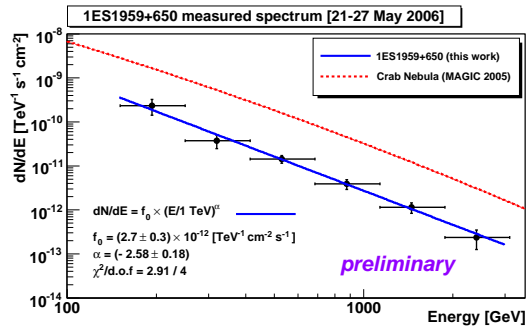}
   \caption{Differential energy distribution of 1ES1959+650 averaged over the whole dataset of the campaign in 2006. The blue solid line represents a power-law fit to the measured spectrum. The fit parameters are listed in the figure. 
  For comparison, the measured MAGIC Crab spectrum~\cite{Crab} is shown by a red dashed line.
   }
   \label{MAGICsp}
   \end{figure}

Figure~\ref{MAGICsp} shows the measured spectrum by the MAGIC telescope. It is well described by a simple power law from 150 GeV to 3 TeV with ${dN/dE} = (2.7 \pm 0.3) \times 10^{-12} \left({E}/{1\ {\rm TeV}}\right)^{-2.58\pm0.18}$ [${\rm TeV}^{-1}\ {\rm s}^{-1}\ {\rm cm}^{-2}$]. Compared to the previous MAGIC measurement of 1ES1959+650 in a steady state in 2004~\cite{MAGIC1959}, the observed flux in 2006 was 60\% of the 2004 result while the photon indices are comparable.

In X-ray, XIS spectra were extracted for $t < 10^5$ s in order to perform a joint XIS and HXD/PIN fit. The model with three power-laws is required to obtain reasonable fitting results. The best fit parameters with associated errors are: $\Gamma _1 = 1.94\pm 0.01, \Gamma _2 = 2.195\pm 0.003, \Gamma _3 = 2.75^{+0.35}_{-0.15}, E_{\rm br,1} = 1.83\pm0.03, E_{\rm br, 2} = 16^{+4}_{-2}$ with $\chi^2/d.o.f = 4065.0/4139$.   

  \begin{figure}[t]
  \centering
 \includegraphics[width=7cm]{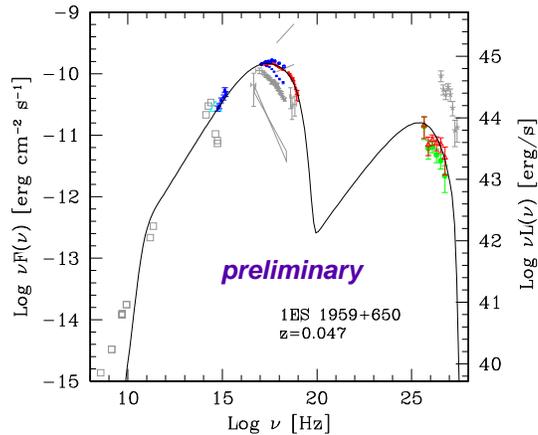}
   \caption{Overall SED of 1ES1959+650. Optical-UV data are from REM (cyan triangle) and UVOT/Swift (blue triangles). We report the average Suzaku spectrum (red) and the swift spectra taken on May 24 [higher] and May 29 [lower]. Green points (filled circles) report the observed MAGIC spectrum, while the red points (empty triangles) have been corrected
for the absorption. Historical data are taken from~\cite{Tag03} (radio-optical-X-rays), \cite{Bec02, Kra04} (X-rays), and~\cite{Aha03} (TeV). The line refers to the best fit SSC model for the campaign data in 2006 using the code developed by~\cite{Tav98,Tav01}. See text for the fit parameters.
   }
   \label{SSC}
   \end{figure}

%

\subsection{Spectral Energy Distributions}

In figure~\ref{SSC} we report the spectral energy distribution (SED) of 1ES1959+650 as measured at the end of
May 2006, together with other historical data~[see the figure caption for the detailed information]. The absorption of VHE $\gamma$-ray emission by the extra-galactic background light (EBL) are corrected using "Low model" of~\cite{Kne04}. \\
Assuming the uniform injection of the electrons throughout a homogeneous emission region 
we model the observed SED by using a one-zone SSC model developed by~\cite{Tav98, Tav01}.
Briefly, a spherical shape (blob) is adopted for the emission region with a radius $R$, filled with a tangled magnetic field with intensity $B$ .
The emission region is moving at a relativistic speed $\beta c$ with a bulk Lorentz factor $\Gamma$ and the viewing angle of $\theta$. The beaming (Doppler) factor is given by $\delta = 1/[\Gamma(1-\beta \cos \theta)]$, with $\delta \sim \Gamma$ for $\theta \sim 1/\Gamma $.
The electron distribution is described by a smoothed broken power-law $N(\gamma) = K\gamma^{-s_1}(1+\gamma/\gamma_{\rm b})^{s_1-s_2} \exp(-\gamma/\gamma _{\rm max})$ where $K$ is a normalization factor,  $\gamma _{\rm b}$ is the break Lorentz factor of electrons, and $s_1$ and $s_2$ are the spectral indices below and above the break, respectively. 
The best fit SSC model was achieved using the following parameters: 
$\delta=35$, $R=5\times 10^{15}$ cm, $B=0.12$ G, $K=2\times 10^{3} {\rm cm}^{-3}$, $\gamma _{\rm min} =1$, $\gamma _{\rm b}=8\times 10^4$, $\gamma _{\rm max} =7\times 10^5$, $s_1=2$ and $s_2=3.5$. 

\section{Summary}
For the first time, in May 2006, we successfully performed simultaneous multiwavelength observations in the VHE $\gamma$-ray, X-ray and optical-UV bands for 1ES1959+650 in a quiescent state. 
The source was clearly detected in all observed energy bands.
The results indicate that the activity of the source was rather stable. 
The VHE $\gamma$-ray flux showed a low state of activity with about 10\% Crab flux unit above 300 GeV and a photon index of $-2.58\pm0.18$.
In the X-ray band the flux was relatively high and only small variability ($\sim$ 10\%) can be seen.
The SED with the simultaneous spectral data in the steady state can be well described by the one-zone SSC model. 
More details will be available in~\cite{Tag07}.

\section{Acknowledgements}
The MAGIC collaboration thanks the IAC for the excellent working conditions at the ORM. The MAGIC project is supported by German BMBF and MPG, the Italian INFN, the Spanish CICYT, the Swiss ETH and the Polish MNiI.

\bibliography{icrc0911}
\bibliographystyle{plain}

\end{document}